\documentclass[12pt]{spieman}  
\usepackage{amsmath,amsfonts,amssymb}
\usepackage{graphicx}
\usepackage{setspace}
\usepackage{subfigure}
\usepackage[subfigure]{tocloft}
\usepackage{hyperref}
\usepackage{multirow}
\usepackage{booktabs,array}
\usepackage{makecell}

\usepackage{amssymb}
\usepackage{amsmath}
\usepackage{comment}
\usepackage[table]{xcolor}

\title{Selecting the Best Optimizers for Deep Learning based Medical Image Segmentation}

\author[a]{Aliasghar Mortazi}
\author[b]{Vedat Cicek}
\author[c]{Elif Keles}
\author[c]{Ulas Bagci}
\affil[a]{Volastra Therapeutics, NYC, NY, USA.}
\affil[b]{Department of Cardiology, Health Sciences University, Istanbul, Turkey.}
\affil[c]{ Machine \& Hybrid Intelligence Lab, Department of Radiology, Northwestern University, Chicago, IL, USA.}

\cftpagenumbersoff{figure}
\cftpagenumbersoff{table} 
\begin{document} 
\maketitle

\begin{abstract}

\noindent\textbf{Purpose:} The goal of this work is to identify the best optimizers for deep learning in the context of cardiac image segmentation and to provide guidance on how to design segmentation networks with effective optimization strategies. \\
\textbf{Approach:} Most successful deep learning networks are trained using two types of stochastic gradient descent (SGD) algorithms: adaptive learning and accelerated schemes. Adaptive learning helps with fast convergence by starting with a larger learning rate (LR) and gradually decreasing it. Momentum optimizers are particularly effective at quickly optimizing neural networks within the accelerated schemes category. By revealing the potential interplay between these two types of algorithms (LR and momentum optimizers or momentum rate (MR) in short), in this article, we explore the two variants of SGD algorithms in a single setting. We suggest using cyclic learning as the base optimizer and integrating optimal values of learning rate and momentum rate. The new optimization function proposed in this work is based on the Nesterov accelerated gradient optimizer, which is more efficient computationally and has better generalization capabilities compared to other adaptive optimizers.\\
\textbf{Results:} We investigated the relationship of LR and MR under an important problem of medical image segmentation of cardiac structures from MRI and CT scans.  We conducted experiments using the cardiac imaging dataset from the ACDC challenge of MICCAI 2017, and four different architectures shown to be successful for cardiac image segmentation problems. Our comprehensive evaluations demonstrated that the proposed optimizer achieved better results (over a 2\% improvement in the dice metric) than other optimizers in deep learning literature with similar or lower computational cost in both single and multi-object segmentation settings.\\ 
\textbf{Conclusions:}  We hypothesized that combination of accelerated and adaptive optimization methods can have a drastic effect in medical image segmentation performances. To this end, we proposed a new cyclic optimization method (\textit{CLMR}) to address the efficiency and accuracy problems in deep learning based medical image segmentation. The proposed strategy yielded better generalization in comparison to adaptive optimizers.
\end{abstract}


{\noindent \footnotesize\textbf{*}Ulas Bagci,  \linkable{ulasbagci@gmail.com} }

\begin{spacing}{2}   

\section{Introduction}
\label{sect:intro}  
Optimization algorithms are used in the training phase of deep learning, where the model is presented with a batch of data, the gradients are calculated, and the weights and biases are updated using an optimization algorithm. Once the model has been trained, it can then be used for inference on new data.

Stochastic gradient descent (SGD) algorithms are the main optimization techniques used to train deep neural networks. These algorithms can be divided into two categories: adaptive learning rate methods (e.g., Adam and AdaGrad) and accelerated schemes (e.g., Nesterov momentum). Both the learning rate (LR) and momentum rate (MR) are important factors in the optimization process. LR, in particular, is a key adjustable parameter that has been extensively studied and modified over the years. The momentum term was introduced to the optimization equation by Rumelhart and Williams in 1986 to allow for larger changes in the network weights without causing oscillation \cite{Rumelhart}. 

There have been controversial results in the literature about the characteristics of available optimization methods. Therefore, there is a need for exploring which optimization method should be chosen for particular tasks. Most neural network optimizers have been evaluated and tested on classification tasks, which have much lower output dimensions compared to segmentation tasks, which have much higher output dimensions. Hence, these differences between classification and segmentation problems imply a different investigation and method for optimization. In this paper, we develop a new optimization method by exploring LR and MR  optimizers for medical image segmentation problems for the first time in the literature. Our proposed  optimizer is simple and promising because it fixes the problems with traditional optimizers and demonstrates how a simple new formulation can solve surprisingly these problems.

\textbf{Non-adaptive vs. adaptive optimizers:} SGD is the dominant optimization algorithm in deep learning,  which is simple and performs well across many applications. However, it has the disadvantage of scaling the gradient \textit{uniformly} in all directions (for each parameter of network). Another challenge in SGD is to choose an appropriate value for LR. Since LR is a  fixed value in SGD based approaches, it is critical to set it up appropriately since it can directly affect both the  convergence speed and prediction accuracy of neural networks. There have been several studies trying to solve this problem by adaptively changing the LR during training which are mostly known as "adaptive optimizers". Based on the history of changes in gradients during network optimization, LR is adapted in each iteration. Examples of such methods consist of \textit{ADAM} \cite{adam}, ADAGrad \cite{adagrad}, and RMSProp \cite{rmsprop}.  In general, adaptive optimizers make training  faster, which has led to their wide use in deep learning applications. 

The development of momentum in neural network optimizers has followed a similar trajectory as the learning rate. Momentum optimizers \cite{momentum} introduced to speed up convergence by considering the changes from last iteration with a multiplier, which called \textit{momentum}, in updating parameters in current iteration. Selecting an appropriate value for the momentum rate (MR) was initially difficult, but this issue was addressed with the introduction of adaptive optimizers like ADAM, which can adaptively adjust both the MR and LR. These adaptive optimizers have become very popular in the field because they quickly converge on training data.

Although they are widely used, adaptive optimizers may converge to different local minima compared to classical SGD approaches, which can lead to worse generalization and out-of-sample performance. This has been demonstrated by a growing number of recent studies \cite{marginal, clr, yellowfin}. To improve the generalization ability of neural networks, researchers have returned to using original SGD approaches but with new strategies for improving convergence speed. For example, the \textit{YellowFin} optimizer demonstrated that manually tuning the learning rate and momentum rate can lead to better results than using the \textit{ADAM} optimizer~\cite{yellowfin}. Although it was a proof-of-concept study that provided evidence for the counterintuitive idea that non-adaptive methods can be effective. However, in practical applications, manually tuning these rates is challenging and time-consuming.

In another attempt, a cyclic learning rate (CLR) was introduced in \cite{clr} to change the LR  according to a cycle (i.e triangle or Gaussian), proposing a practical solution to hand-tuning requirements. The CLR's only disadvantage was that a fixed MR could limit the search states of LR and MR and cause them to fail until find an optimal solution. Our work will go beyond this constraint. 

\textbf{Summary of our contribution: }
By motivated from~\cite{clr}, herein  we introduce a new version of CLR called "Cyclic Learning/Momentum Rate" (CLMR). This new optimizer alternates the values of the learning rate and momentum rate during training, which has two benefits compared to adaptive optimizers. First, it is more efficient computationally. Second, it has better generalization performance. Furthermore, \textit{CLMR} leads to better results than conventional approaches such as SGD and CLR. Lastly, we investigate the effect of changing the frequency of cyclic function in training and generalization and suggest the optimum frequency values. We investigate several optimizers commonly used in medical image segmentation problems, and compare their performance as well as generalization ability in single and multi-object segmentation settings by using cardiac MR images (Cine-MRI).   

The rest of the paper is organized as follows. In Section 2, we introduce the background information for neural network optimizers, their notations and their use in medical image segmentation. In section 3, we give the details of the proposed method and network architectures on which segmentation experiments have been conducted. Experimental results are summarized in Section 4. Section 5 concludes the paper with discussions and future work.

\section{Background}
 
Optimizing a deep neural network, which is a high-dimensional system with millions of parameters, is one of the most challenging aspects of making these systems more practical. Designing and implementing the best optimizer for deep network training has received a lot of attention in recent decades. These studies mainly address two major issues: (1) Making the network training as fast as possible (fast convergence), (2) increasing the  generalizability of networks.  SGD optimizers have been  the most popular optimizer in deep networks due to their low computational cost, fast convergence. There have been major modifications to original SGD optimizer during last decade to make them for efficient for training deep nets. The following are some of the key optimization studies related to our efforts.      

\subsection{Optimizers with fixed LR/MR}
SGD and Mini-batch gradient descent were first optimizers used for training neural networks. The updating rule for these optimizers include only the value of last iteration as shown in Eq. \ref{gd_update}. Choosing appropriate value for a LR  is challenging in these optimizers since if LR is very small then convergence is very slow; and if LR is set high, the optimizer will oscillate around global minima instead of converging: 

\begin{equation}\label{gd_update}
~\theta_i = ~\theta_{i-1}- ~\alpha ~\nabla_{~\theta_i}J(~\theta_i),
\end{equation}
where $\theta$ is network parameters, $~\alpha$ is LR, \textit{J} is cost function to be minimized (function of $\theta$, $X$(input), and $Y$(labels)). The equation \ref{gd_update} can be considered as an updating rule for SGD and mini-batch gradient descent by choosing $X$ and $Y$ as whole samples, a single sample, or a batch of samples  in a dataset.  

The Momentum optimizer was designed to accelerate the optimization process by taking into account the values from previous iterations, weighted by a factor known as "momentum," as mentioned in \cite{momentum}. The updating for this optimizer is defined as: 
\begin{equation}\label{mom_update}
~\theta_i = ~\theta_{i-1}- ~\alpha ~\nabla_{~\theta_i}J(~\theta_i)- ~\beta(~\theta_{i-1}-~\theta_{i-2}),
\end{equation}
where $~\beta$ denotes the \textit{momentum rate} (MR). In \textit{Momentum} optimizer, the past iterations don't play any role in cost function and cost function is only calculated for the current iteration only. Also, similar to LR, choosing a proper value for MR is challenging and it has a correlation with LR too. 

\textit{Nesterov accelerated gradient}~\cite{nag} (NAG) was then introduced  to address the limitation of momentum optimizers as well as to accelerate the convergence by including information from previous iterations in calculating the gradient of the  cost function as shown in the following equation: 
\begin{equation}\label{nag_update}
~\theta_i = ~\theta_{i-1}- ~\alpha ~\nabla_{~\theta_i}J(~\theta_i- ~\beta(~\theta_{i-1}-~\theta_{i-2}))- ~\beta(~\theta_{i-1}-~\theta_{i-2}).
\end{equation}
Compared to optimizers with fixed LR/MR, the NAG optimizer generally shows improved performance in both convergence speed and generalizability.

\subsection{Optimizers with adaptive LR and MR}
A significant disadvantage of optimizers with a fixed LR/MR is that they cannot incorporate information from the gradients of past iterations in adjusting the learning and momentum rates. For instance, they cannot increase the learning rate for dimensions with a small slope to improve convergence, or reduce the learning rate for dimensions with a steep slope to avoid oscillation around the minimum point. \textit{Adagrad}~\cite{adagrad} is one of fist adaptive LR optimizers used in deep networks adapting the learning rate for each parameter in the network by dividing the gradient of each parameter by its sum of the squares of gradient, as follows: 
\begin{equation}\label{adagrad_update}
~\theta_i = ~\theta_{i-1}- ~\alpha ~\frac{1}{~\sqrt[]{G_i+~\epsilon}}~\circ ~\nabla_{~\theta_i}J(~\theta_i),
\end{equation}
where $G_i$ is a diagonal (square) matrix and each diagonal element equal to the sum of the square of gradient of its corresponding parameters: 
\begin{equation}\label{g_matrix}
G_i = \displaystyle\sum_{i=1}^I(~\nabla_{~\theta_i}J(~\theta_i))^2,
\end{equation}
where ~\textit{I} is the current iteration. 

One of the drawbacks of~\textit{AdaGrad} is gradient vanishing due to accumulation of all past square gradient in denominator of Equation~\ref{adagrad_update} during the training. This leads the gradients to converge to zero after several epochs in training. However, ~\textit{AdaDelta}, ~\textit{RMSProp}, and ~\textit{ADAM} optimizers solved this problem by considering a sum of the past samples within a pre-defined window. \textit{ADAM} optimizer's updating rule uses past squared gradient (as scale) and also like momentum, it keeps  an exponentially decaying average of past gradients. Hence, these adaptive optimizers have advantages over the \textit{AdaGrad} by adaptively changing both RL and ML as well as resolving the gradient vanishing issue: 
\begin{equation}\label{adagrad_update}
~\theta_i = ~\theta_{i-1}- ~\alpha_i ~\frac{~\beta_1 ~\nabla_{~\theta} J(~\theta_{i-2})-(1-~\beta_1)~\nabla_{~\theta}J(~\theta_{i-1})}{~\sqrt[]{~\beta_2+~\epsilon}} ~\circ ~\nabla_{~\theta_i}J(~\theta_i).
\end{equation}

Adaptive learning methods are costly because they are required to calculate and keep all the past gradients and their squares to update the next parameters. Also, the adaptive learning optimizer may converge  into different minima  in comparison with fixed learning rate optimizers~\cite{clr, marginal, yellowfin}. 

Alternatively, Cyclic learning rate (CLR) was proposed to change the learning rate during training, which needed no additional computational cost.  CLR is a method for training neural networks that involves periodically changing the learning rate during training. As mentioned earlier, the learning rate is typically adjusted according to a predetermined schedule, such as increasing the learning rate from a low value to a high value and then decreasing it back to the low value over a set number of training iterations. The learning rate is then reset and the process is repeated. This can help the optimization process by allowing the model to make larger updates at the beginning of training and smaller updates as training progresses, potentially leading to faster convergence and better model performance~\cite{clr}. Later in Figure~\ref{fig:C}a, we show how we use CLR in our methodology.

\subsection{Cardiac Image Segmentation}
Cardiovascular diseases (CVDs) are the leading cause of death worldwide according to the World Health Organization (WHO). CVDs lead to millions of deaths annually and are expected to cause over 23.6 million deaths in 2030~\cite{who}. Cine-MR imaging can provide valuable information about cardiac diseases due to its excellent soft tissue contrast. For example, ejection fraction (EF), an important metric measuring how much blood the left ventricle pumps out with each contraction, can be measured with Cine-MRI. To this end, radiologists often manually measure the volume of the heart at the end of the systole (ES) and the end of the diastole (ED) to measure EF. This is a time-consuming process with known inter-, and intra-observer variations. Due to its significance in functional assessment of heart, there have been numerous machine learning based automated algorithms developed in the literature for measuring EF. In this study, we make our efforts in this application due to its importance in the clinic. 

There is a considerable amount of research dedicated to the problem of cardiac segmentation from MR or CT images. Since Xu et al. found a correlation between motion characteristics and tissue properties, they developed a combined motion feature learning architecture for distinguishing myocardial infarction \cite{xu2018direct}. In our another attempt, \textit{CardiacNet} in \cite{cardiacnet} proposed a multi-view CNN to segment the  left atrium and proximal pulmonary veins from MR images following by an adaptive fusion. The shape prior information from deep networks were used to guide segmentation network to delineate cardiac substructures from MR images \cite{mortazi2019weakly, oktay2017anatomically}. As previously stated, the literature and methodologies for cardiac segmentation are extensive. Readers are invited to consult references \cite{bizopoulos2018deep} and \cite{zhuang2019evaluation} for more comprehensive information. 

\section{Methods}
We approach the optimization problem from the perspective of a significant medical image analysis application: segmentation. Segmentation is rarely studied from an optimization perspective in comparison to classification.

\begin{figure}[t]
\centering
\includegraphics[width=1\textwidth]{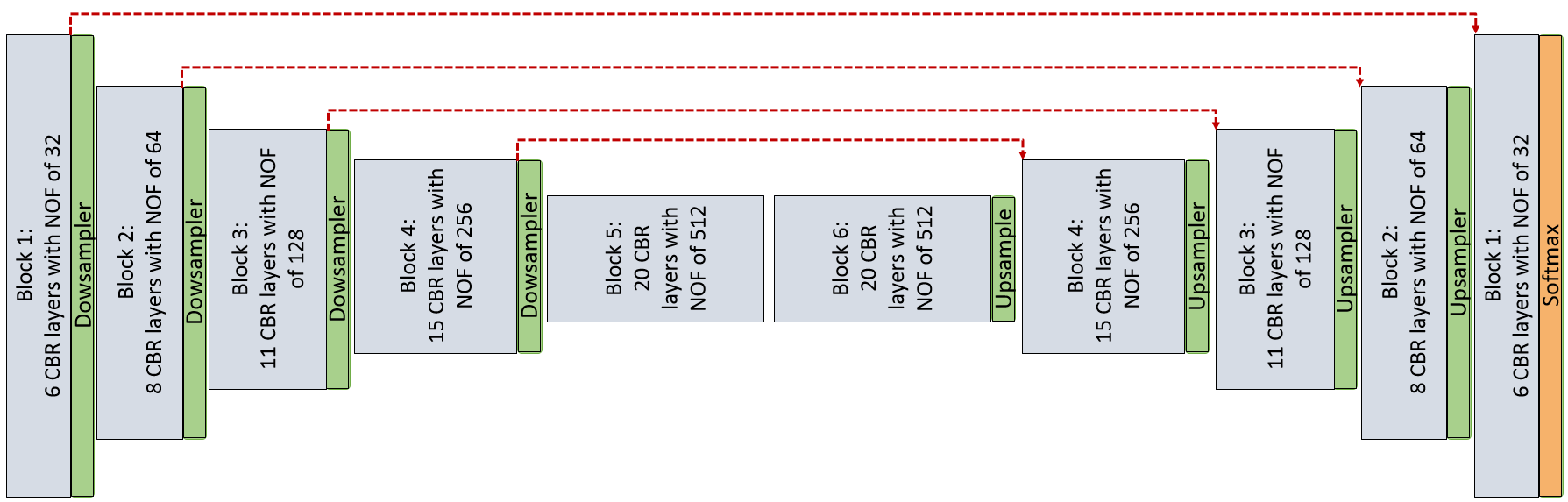}
\caption{CNN Architecture is used for pixel-wise segmentation. The architecture with CNN blocks  without red skip connections is Encoder-Decoder architecture. The architecture with red skip connection  (Fig. \ref{fig:opt_Bl}a) is called \textit{U-Net}, if connections are with Dense block (Fig. \ref{fig:opt_Bl}b), it is called Tiramisu (DenseNet for segmentation) \label{fig:opt_arch}}
\end{figure}

Over the past few years, there has been a dramatic increase in the use of CNN in computer vision and medical imaging applications, more recently combined with Transformers~\cite{u1, u2, u3, u4, u5}. The successful CNN-based segmentation approaches can be divided into three  broad categories. The first category is named encoder-decoder architecture. One of the most famous works in this category has been done by Badrinarayanan and et al.~\cite{segnet}, called \textit{SegNet}, and it is designed for semantic segmentation. Another category of architecture is called \textit{ResNet}~\cite{resnet} and it is proposed  by He and et al. for image recognition but later, the \textit{U-Net} \cite{unet} was proposed from the similar extending recognition into segmentation with a U-shaped network consisting of skip connections between encoder and decoder. The last category of the architecture is based on the \textit{DenseNet}\cite{denselyCNN}, instead of having a residual connection, the vectors are concatenated to each other to  maximize the information flow through the network. Hence, the information loss during backpropagation can be minimized by considering these connections. \textit{DenseNet} itself is proposed for image classification, but by combining concepts of the \textit{ResNet} and \textit{DenseNet}, a new architecture was introduced in \cite{tiramisu} in a U-Net shape to do segmentation. There are many more architectures based on the U-Net style with adaptation from the CNN and Transformers literature. In our study, we conducted experiments in three different (mostly used) segmentation architectures to demonstrate the effect of the connections, as explained in the following subsection. One may increase the number of architectures for more comparisons, but this is outside the scope of our study. CNN Architectures used in the experiments are the following:



\textbf{1. Encoder-Decoder Architecture}: This architecture simply consists of the encoder and decoder part as illustrated in Figure~\ref{fig:opt_arch}, \textbf{without} considering red skip connections. The filter size in all the layer are $3\times 3$ and each encoder and decoder part include $5$ \textit{CNN blocks} and each~\textit{CNN blocks} consist of different number of layers as mentioned in Table \ref{tab:arch}. Also, the number of filters in each~\textit{CNN block} are a fixed number and they are mentioned in Table~\ref{tab:arch} for each layer. Each layer within the \textit{CNN block} includes \textit{Convolution+Batch normalization+ReLu} as activation function (\textit{CBR}).

\textbf{2. U-Net Architecture}: U-Net is particularly popular in medical image analysis. The U-Net model is based on a fully convolutional network, which means that it is built entirely out of convolutional layers and does not contain any fully connected layers. This makes it well-suited for image segmentation tasks, as it can process input images of any size and output a corresponding segmentation map. The U-Net model is known for its ability to handle small, sparsely annotated training datasets, making it a useful tool for medical image analysis where such datasets are common. This architecture is similar to the Encoder-Decoder architecture as illustrated in Figure \ref{fig:opt_arch} \textbf{with} red skip connections from encoder  to decoder. The number of layers and filters for each block are mentioned in Table \ref{tab:arch}.  

\textbf{3. DenseNet Architecture}: DenseNet is another convolutional neural network architecture that was developed to improve upon the efficiency of training deep networks. The key idea behind DenseNet is to connect all layers in the network directly to every other layer, rather than only connecting each layer to its immediate neighbors as is done in traditional convolutional networks. This allows the network to learn more efficient feature representations and reduces the risk of overfitting. DenseNets have been successful in a number of applications and have achieved state-of-the-art performance on image classification and segmentation tasks. We will use two different \textit{DenseNet} architectures in our experiments. First, the architecture in Figure~\ref{fig:opt_arch} with dense blocks (DBs) and skip connections is \textit{DenseNet}\_1. Then, in order to use higher growth rate (GR), in \textit{DenseNet}\_2, at the end of each block a convolution layer with kernel size of $1\times1$ is used to decrease number of its input filters by \textbf{C} rate, which \textbf{C} is equal 2 in this paper. The GR in \textit{DenseNet}\_2 increased to 24 (from 16 in \textit{DenseNet}\_1) while the number of parameters decreased (Table \ref{tab:arch}).The number of CBR layers and also the  number of parameters are mentioned in Table \ref{tab:arch}. 

\begin{table}[h]
\small
\centering
\caption{Number of layers in each block of different architectures and number of parameters.\label{tab:arch}}
\resizebox{\columnwidth}{!}{
\begin{tabular}{c|c|c|c|c}
\hline
 & \textbf{Enc\_Dec} & \textbf{U-Net} & \textbf{\begin{tabular}[c]{@{}c@{}}DenseNet\_1\\  (GR=16)\end{tabular}} & \textbf{\begin{tabular}[c]{@{}c@{}}DenseNet\_2\\ (GR=24)\end{tabular}} \\ \hline
\textbf{Block 1} & \begin{tabular}[c]{@{}c@{}}6 layers,\#filters=32\end{tabular} & \begin{tabular}[c]{@{}c@{}}6 layers,\#filters=32\end{tabular} & 6 layers & 6 layers \\ \hline
\textbf{Block 2} & \begin{tabular}[c]{@{}c@{}}8 layers,\#filters=64\end{tabular} & \begin{tabular}[c]{@{}c@{}}8 layers,\#filters=64\end{tabular} & 8 layers & 8 layers \\ \hline
\textbf{Block 3} & \begin{tabular}[c]{@{}c@{}}11 layers,\#filters=128\end{tabular} & \begin{tabular}[c]{@{}c@{}}11 layers,\#filters=128\end{tabular} & 11 layers, & 11 layers \\ \hline
\textbf{Block 4} & \begin{tabular}[c]{@{}c@{}}15 layers,\#filters=256\end{tabular} & \begin{tabular}[c]{@{}c@{}}15 layers,\#filters=256\end{tabular} & 15 layers & 15 layers \\ \hline
\textbf{Block 5} & \begin{tabular}[c]{@{}c@{}}20 layers,\#filters=512\end{tabular} & \begin{tabular}[c]{@{}c@{}}20 layers,\#filters=512\end{tabular} & 20 layers & 20 layers \\ \hline
\textbf{Block 6} & \begin{tabular}[c]{@{}c@{}}20 layers,\#filters=512\end{tabular} & \begin{tabular}[c]{@{}c@{}}20 layers,\#filters=512\end{tabular} & 20 layers & 20 layers \\ \hline
\textbf{Block 7} & \begin{tabular}[c]{@{}c@{}}15 layers,\#filters=256\end{tabular} & \begin{tabular}[c]{@{}c@{}}15 layers,\#filters=256\end{tabular} & 15 layers & 15 layers \\ \hline
\textbf{Block 8} & \begin{tabular}[c]{@{}c@{}}11 layers,\#filters=128\end{tabular} & \begin{tabular}[c]{@{}c@{}}11 layers,\#filters=128\end{tabular} & 11 layers & 11 layers \\ \hline
\textbf{Block 9} & \begin{tabular}[c]{@{}c@{}}8 layers,\#filters=64\end{tabular} & \begin{tabular}[c]{@{}c@{}}8 layers,\#filters=64\end{tabular} & 8 layers & 8 layers \\ \hline
\textbf{Block 10} & \begin{tabular}[c]{@{}c@{}}6 layers,\#filters=32\end{tabular} & \begin{tabular}[c]{@{}c@{}}6 layers,\#filters=32\end{tabular} & 6 layers & 6 layers \\ \hline
\textbf{\begin{tabular}[c]{@{}c@{}}\# of params\\  (in million):\end{tabular}} & 77.5 & 79.1 & 7.7 & 8.8 \\ \hline
\end{tabular}
}
\end{table}

\subsection{Dense Block}
Within the DB, a concatenation operation is done for combining the feature maps (through direction (axis) of the channels) for the last three layers. So, if the input to $\it{l^{th}}$ layer is $\mathbf{X}_{\it{l}}$, then the output of $\it{l^{th}}$ layer is:
\begin{equation}\label{layer_eq}
 	F(\mathbf{X}_{\it{l}})= CBR(\mathbf{X}_{\it{l}}). 
 \end{equation}
 Since we are doing concatenation before each layer (except the first one), the output of each layer can be calculated only by considering the input and output of first layer as following: 

\begin{equation}\label{concat}
\begin{split}
 	F(\mathbf{X}_{\it{l}})=F(\displaystyle \underset{{\it{l}'=0}}{\overset{{\it{l}'=\it{l}-1}}{\mathbf{^\frown}}} F(\mathbf{X}_{\it{l}'})) \qquad \text{for} \quad\it{l}\ge1 \quad 
   \\  \text{and} \quad \it{l}=\{1,2,\dots,\it{L},\},
\end{split}
\end{equation}
where $\mathbf{^\frown}$ is concatenation operation. In addition, for initialization $F(\mathbf{X}_{\it{-1}})$ and $F(\mathbf{X}_{\it{0}})$ are considered as \{\} and $\mathbf{X}_{\it{1}}$ respectively which \{\} is an empty set and there are $\it{L}$ layers inside of the block.

Assuming the number of output features for each layer is  $\mathbf{K_{out}}$ (channel out) and the number of input features for first layer is $\mathbf{K_{in_1}}$ (channel in). Then, the feature maps growth (channel out) for second, third, \dots, and $\it{L^{th}}$ layer are $\mathbf{K_{out}}+\mathbf{K_{in_1}}$, 2$\mathbf{K_{out}}+\mathbf{K_{in_1}}$, \dots, and $(\it{L}-1)\mathbf{K_{out}}+\mathbf{K_{in_1}}$ respectively. The growth rate for the DB is the same as fourth layer.

\begin{figure}[t]
\centering
\includegraphics[ width=1 \textwidth]{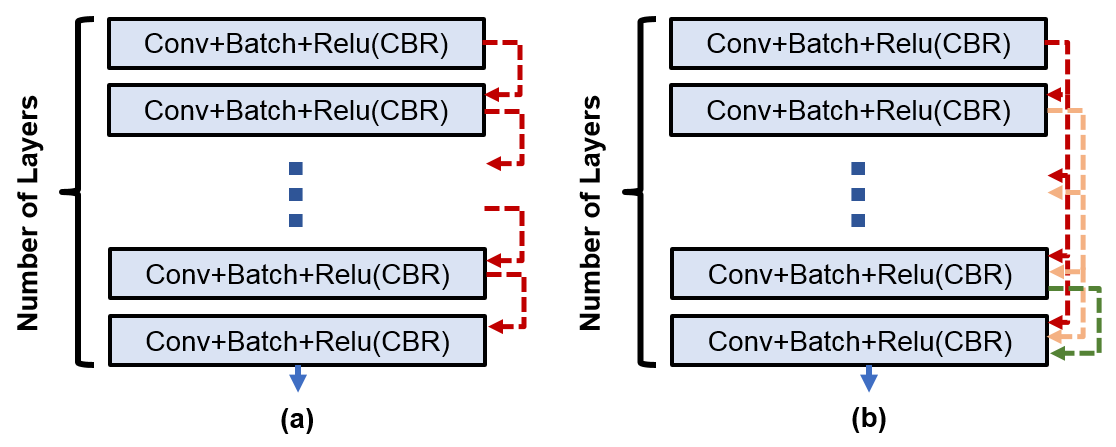}
\caption{(a) CNN block used in Enc-Dec and \textit{U-Net} architectures, (b) Dense block used in Tiramisu architecture.   \label{fig:opt_Bl}}
\end{figure}

\subsection{Cyclic Learning/Momentum Rate Optimizer}
\begin{figure}[h]
\includegraphics[width=0.97\textwidth]{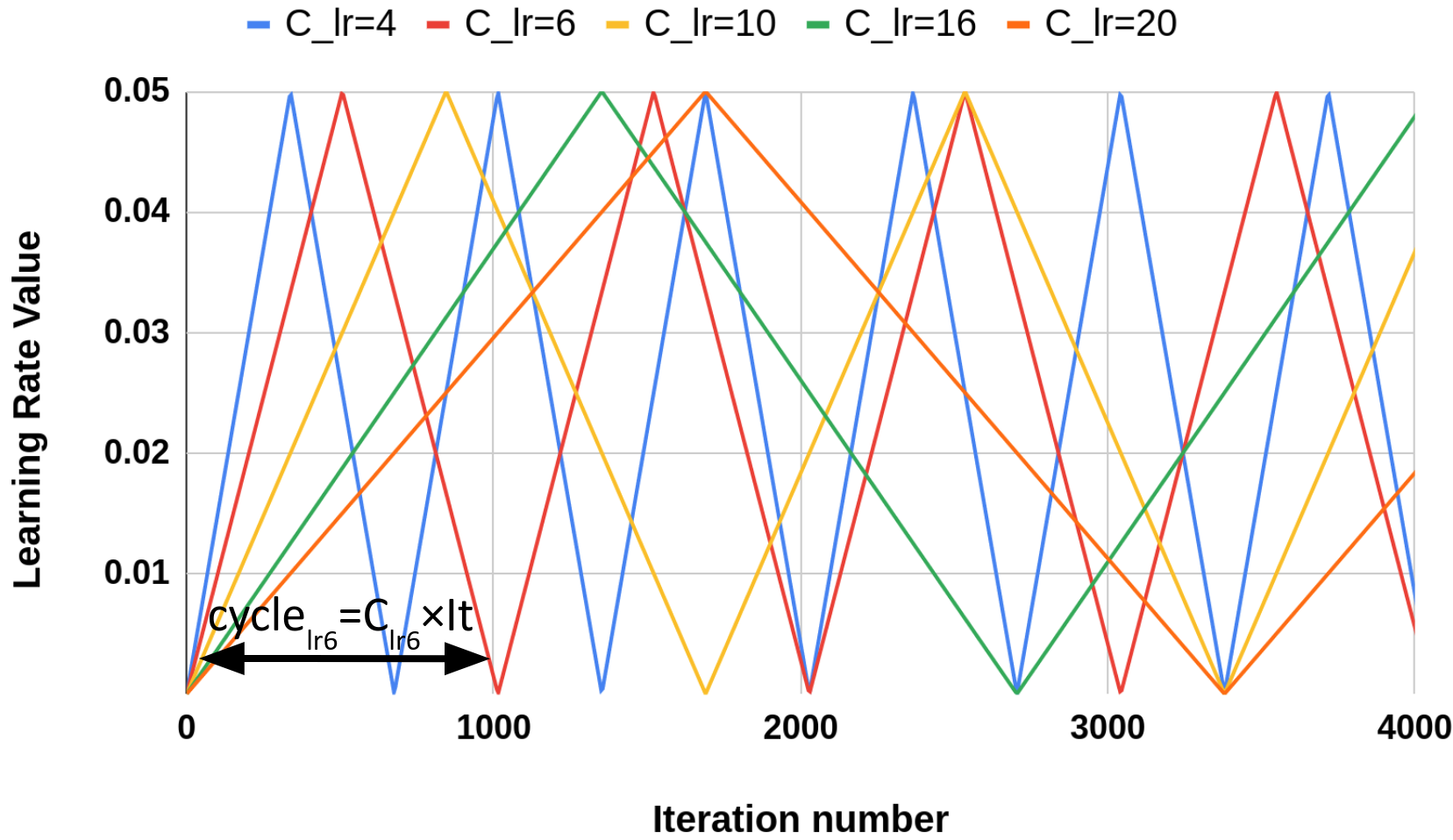}
\includegraphics[width=0.97\textwidth]{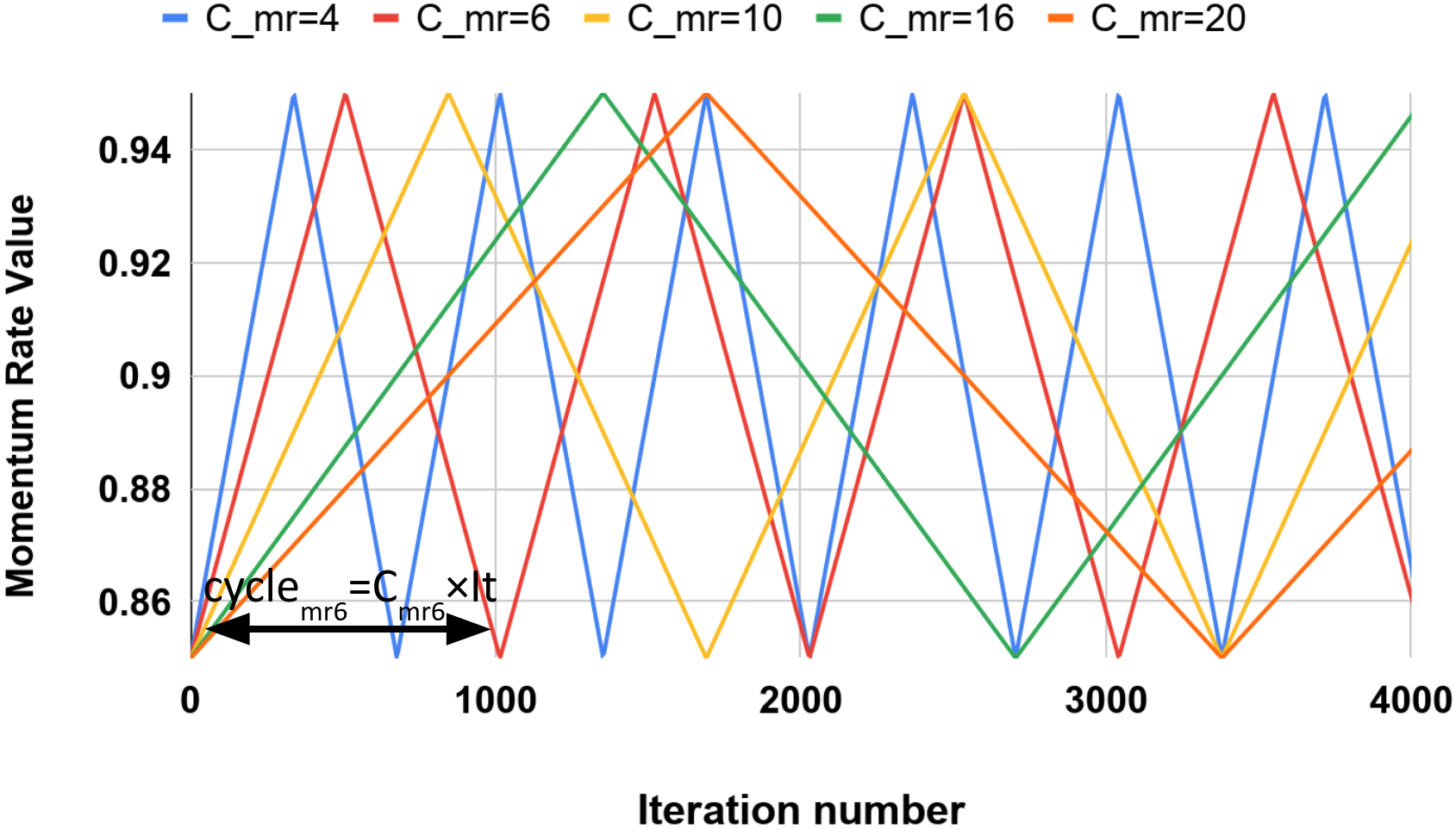}
 \caption{CLR and MLR functions. (a) Learning rate triangle function for different $C_{lr}$ values with $min_{lr}=0.0005$ and $max_{lr}=0.05$. (b) Momentum rate triangle function for different $C_{mr}$ values with $min_{mr}=0.85$ and $max_{mr}=0.95$. \label{fig:C}}
\end{figure}

Smith et al~\cite{clr} argued that a cycling learning may be a more effective alternative to adaptive optimizations especially from generalization perspective. Basically, cyclic learning includes a pre-defined cycle (such as triangle or Gaussian function) that learning rate is changing according to that cycle. Here, we hypothesize (and show later in the results section) that having a cyclic momentum in Nesterov optimizer (Eq. \ref{mom_update}) can lead to a better accuracy in segmentation task in generalization phase. As a reminder, \textit{momentum} in Eq. \ref{mom_update} was used to consider the past iterations by a coefficient called \textit{momentum}. So, choosing the proper value for \textit{momentum} is challenging. To this end, we propose changing the MR in the same way that we changed the LR, and we considered the cyclic triangle function for both MR and LR as  illustrated in the Figure \ref{fig:C}. \textit{$cycle_{lr}$} and  \textit{$cycle_{mr}$} determine the period of triangle function for LR and MR are defined by: 
\begin{equation}\label{cycle}
cycle_{lr}=C_{lr}\times It,
\end{equation}
\begin{equation}\label{cycle}
cycle_{mr}=C_{mr}\times It,
\end{equation}
where \textit{$C_{lr}$} and \textit{$C_{mr}$} are positive even integer numbers, \textit{It} is number of iteration per each epoch. 

In Figures \ref{fig:C}a and \ref{fig:C}b, the cyclic function for different values of \textit{$C_{lr}$} and \textit{$C_{mr}$} are illustrated. LR during whole training can be determined from equation \ref{eq:lr}:  
\begin{equation}
LR = 
\begingroup
\begin{cases}
  2 \times \frac{max_{lr}-min_{lr}}{C_{lr} \times It} \times i + min_{lr},   &  for \quad N \times cycle_{lr}\leq i < \frac{2N+1}{2} \times cycle_{lr}  \\
  
   -2 \times \frac{max_{lr}-min_{lr}}{C_{lr} \times It} \times i + 2max_{lr} - min_{lr},  &  for \quad \frac{2N+1}{2} \times cycle_{lr} \leq i < (N+1) \times cycle_{lr},
\end{cases}
\endgroup
   \label{eq:lr}
\end{equation}
where $max_{lr}$ and $min_{lr}$ are maximum and minimum values of LR function, respectively. $i$ is the iteration indicator during whole training process and $i\in \{ 1, 2, \dots, It\times Ep\}$, which $Ep$ is total number of epochs in training and $N$ is a set of natural number. MR can also be determined as: 
\begin{equation}
MR = 
\begingroup
\begin{cases}
  2 \times \frac{max_{mr}-min_{mr}}{C_{mr} \times It} \times i + min_{mr},   &  for \quad N \times cycle_{mr}\leq i < \frac{2N+1}{2} \times cycle_{mr}  \\
  
   -2 \times \frac{max_{mr}-min_{mr}}{C_{mr} \times It} \times i + 2max_{mr} - min_{mr},  &  for \quad \frac{2N+1}{2} \times cycle_{mr} \leq i < (N+1) \times cycle_{mr},
\end{cases}
\endgroup
   \label{eq:mr}
\end{equation}
where $max_{mr}$ and $min_{mr}$ are maximum and minimum values of MR function, respectively. 

Equations \ref{eq:lr} and \ref{eq:mr} are used to determine the values of LR and MR in each iteration during training. One of the challenges in using these cyclic LR and MR functions are determining the values of some variables in the equations including $max_{lr}$, $min_{lr}$, and $C_{lr}$ for LR; and also $max_{mr}$, $min_{mr}$, and $C_{mr}$ for MR. For finding the the $max_{lr}$ and $min_{lr}$ values, as it suggested in \cite{clr}, one can run the networks with different LR values for a few epochs and then these values are chosen according to how network accuracy changes. Since, when both LR and MR change dynamically and the one value can affect the other one (considering the optimizer formula), it makes more challenging to find the CLMR optimum parameters by proposed solution. It means we need to train large number of networks in order to determine the optimum values of $max_{lr}$, $min_{lr}$, $max_{,r}$, and $min_{mr}$ which is not computationally feasible. Also, a heuristic method was suggested in \cite{clr} to find the best value of $C_{lr}$.          

In this paper we propose an alternative way to find best cyclic functions with minimum computational cost. We set fixed values for $max_{lr}$, $min_{lr}$, $max_{,r}$, and $min_{mr}$ parameters and make sure that the selected values cover a good range of values for both LR and MR in practice (illustrated in Figure \ref{fig:C}). Then, we did a computationally reasonable heuristic search for finding the appropriate amount of $C_{lr}$ and $C_{mr}$ from the values shown in Figure \ref{fig:C}. Since, changing the values of $C_{lr}$ and $C_{mr}$ leads to change in the values of LR and MR in different iterations, there is no need to find the optimum values for minimum and maximum, we did search in 2D space of $C_{lr}$ and $C_{mr}$ to find their optimal values.


\section{Experiments and results}
\subsection{Data}
For investigating the performance of proposed method, a dataset from Automatic Cardiac Diagnosis Challenge (ACDC-MICCAI Workshop 2017) were used \cite{acdc}. These data set includes 150 cine-MR images: 30 normal cases, 30 patients with myocardial infarction, 30 patients with dilated cardiomyopathy, 30 patients with hypertrophic cardiomyopathy, and the remaining 30 patients with abnormal RV. While 100 cine-MR images were used for training (80) and validation (20), the remaining 50 images were used for testing with online evaluation by the challenge organizers. For a fair validation in training procedures, four subjects from each category have been chosen. The binary masks for ground truths of three substructures were provided by the challenge organizers for training and validation while test set was evaluated online (unseen test set. Three substructures are right ventricle (RV), myocardium of left ventricle(Myo.), and left ventricle (LV) at two time points of end-systole (ES)  and end-diastole (ED).    

The MRIs were obtained using two MRI scanners of different magnetic strengths (1.5T and 3.0T). Cine-MR images were acquired with a SSFP sequence in short axis while on breath hold (and gating). In particular, a series of short axis slices cover the LV from the base to the apex, with a thickness of 5 mm (or sometimes 8 mm) and sometimes an inter-slice gap of 5 mm. The spatial resolution goes from 1.37 to 1.68 $mm^2/pixel$ and 28 to 40 volumes cover completely or partially the cardiac cycle.

\subsection{Implementation details}
The networks were trained for  a fixed number of epochs (100) and it was confirmed that they are  fully trained. All the images were resized to $200\times200$ in short axis by using B-spline interpolation. Then, as a preprocessing step, we applied anisotropic filtering and histogram matching to  the whole data set. The total number of 2D slices for training was about 1690 and batch size of 10 were chosen for training. Hence, the number of iteration per epoch is $\frac{1690}{10}=169$ and we have a total number of iteration $100\times169=16,900$ in training. The Cross Entropy loss function was chosen for minimization. All the networks were implemented on Tensorflow with using NVIDIA TitanXP GPUs.  

\subsection{Results }
\begin{figure}[h]
\includegraphics[width=0.90\textwidth]{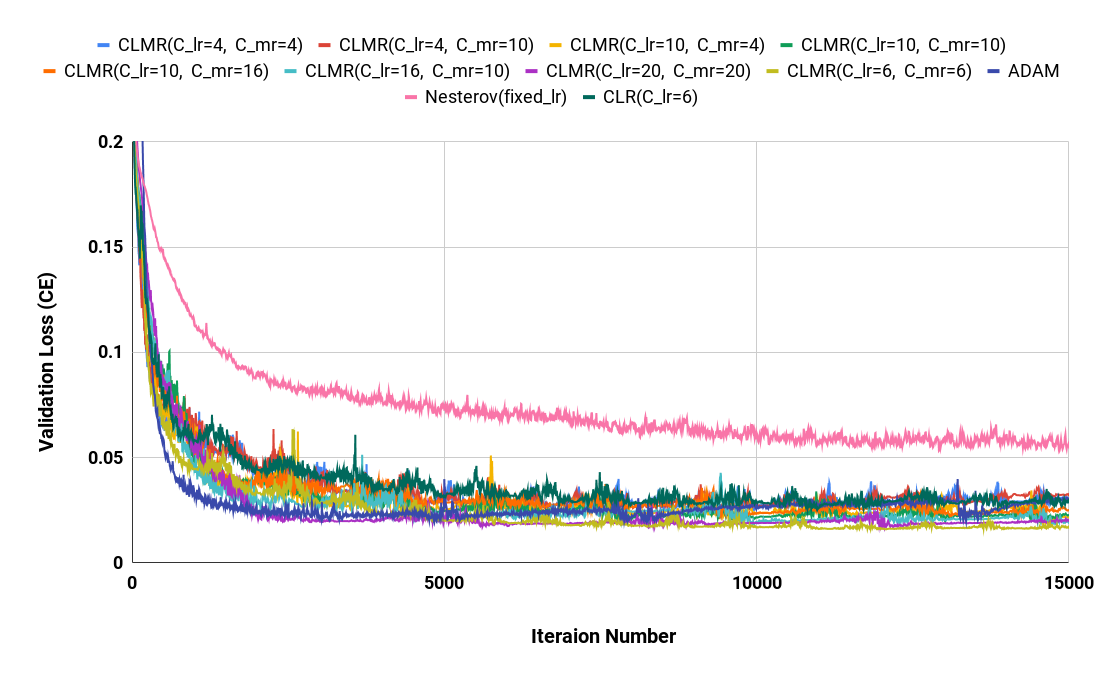}\label{fig:recall_validation}
\includegraphics[width=0.90\textwidth]{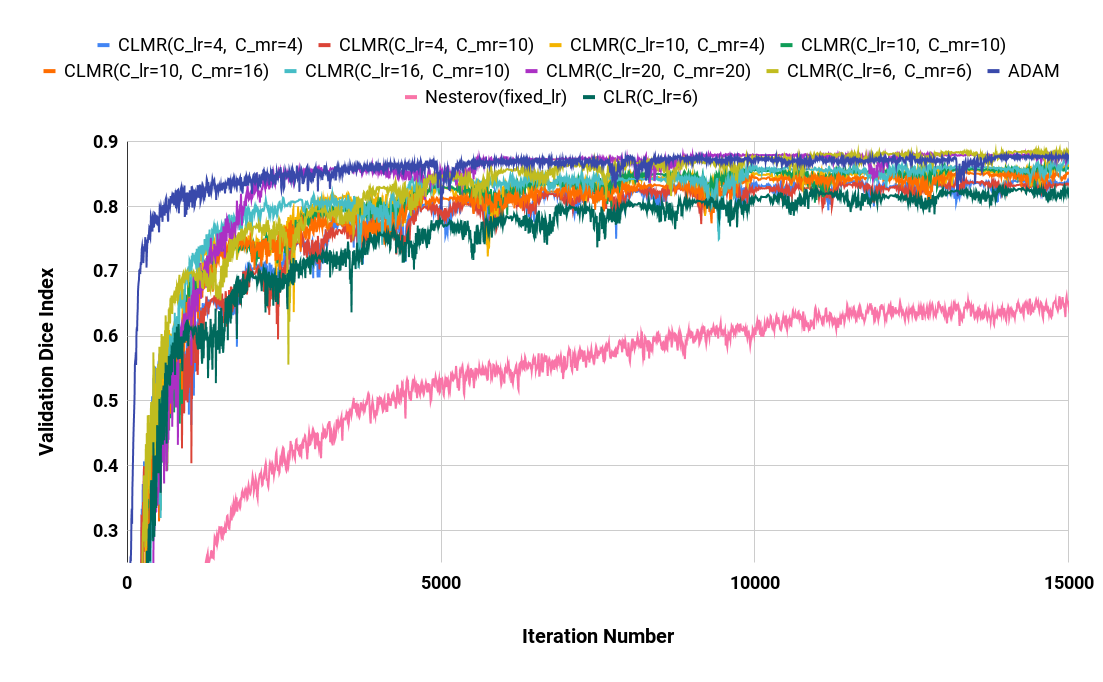}\label{fig:dice_validation}
\caption{Validation loss and dice index for \textit{DenseNet}\_2 architecture with different values of ~\textit{$C_{lr}$} and ~\textit{$C_{mr}$}. (Upper) Cross Entropy loss in validation set for \textit{DenseNet}\_2 architecture. (Lower) Dice index in validation set for \textit{U-Net} architecture (zoomed). \label{fig:unet}}
\end{figure}

\begin{figure}[h]
\includegraphics[width=0.94\textwidth]{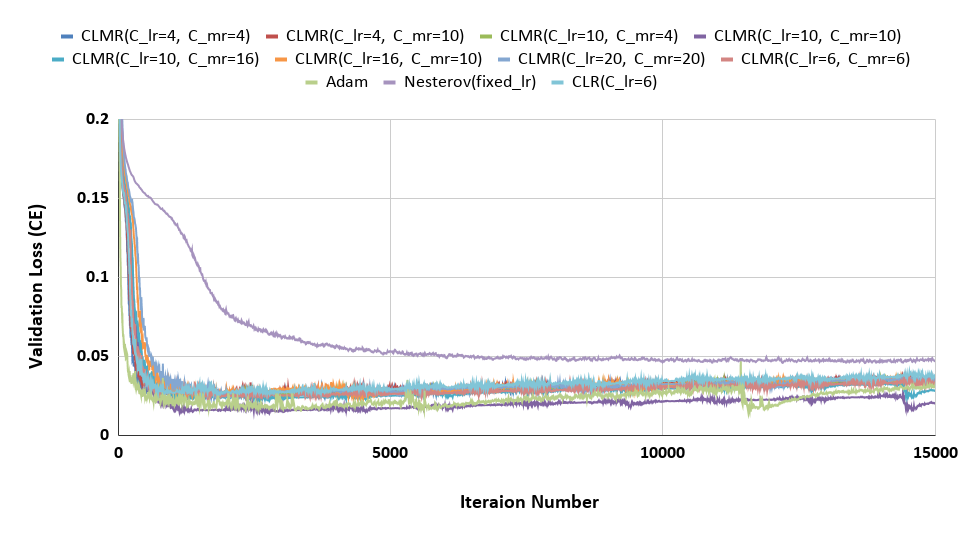}
\includegraphics[width=0.94\textwidth]{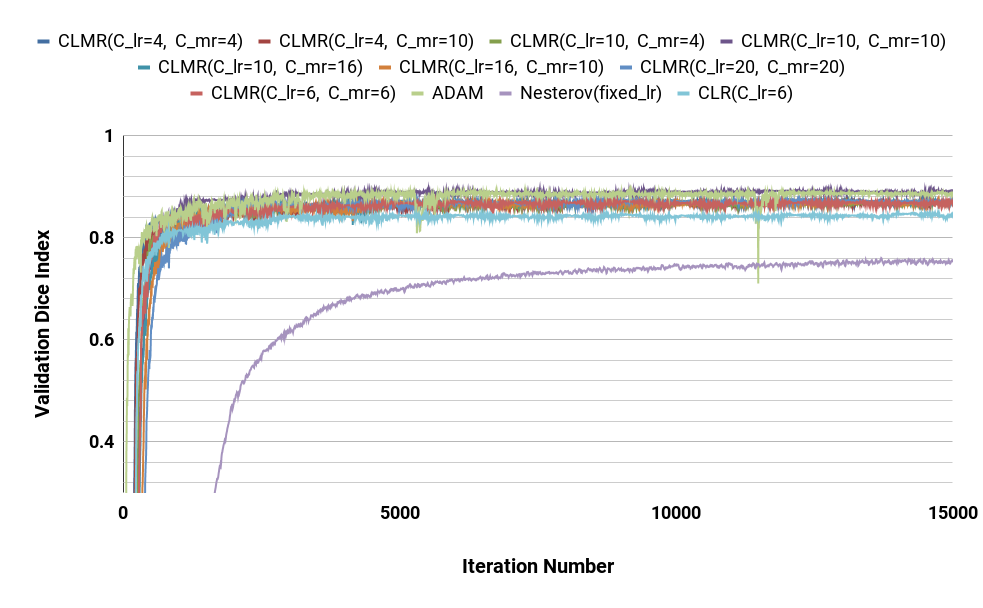}
 \caption{Validation loss and dice index for \textit{U-Net} architecture with different values of ~\textit{$C_{lr}$} and ~\textit{$C_{mr}$}. (Upper) Cross Entropy loss in validation set for \textit{DenseNet}\_2 architecture. (Lower) Dice index in validation set for \textit{U-Net} architecture (zoomed). \label{fig:densenet_2}}
\end{figure}

We calculated Dice Index (DI) and also Cross Entropy (CE) loss on validation set for investigating our proposed optimizer along with other optimizers. In Figures \ref{fig:unet}a and b, the CE and DI curves versus iterations for \textit{U-Net} architecture for different optimizers are illustrated. As these curves  show, the DI in \textit{U-Net} with \textit{ADAM} optimizer is increasing rapidly and sharply at the very beginning and then it is almost fixed afterwards. Although our proposed optimizer (CLMR(C\_lr=20, C\_mr=20)) is not learning as fast as \textit{ADAM} optimizer at very beginning in \textit{U-Net}, it gets better accuracy than \textit{ADAM} finally. This phenomenon is clearer in CE curves. The quantitative results on test set in Table \ref{tab:results} support the same observation and conclusion. Further, the same pattern happens for \textit{DenseNet}\_2 architecture in Figure \ref{fig:densenet_2}. This confirms our hypothesis that adaptive optimizers converges faster but to potentially to different local minimas in comparison with classical SGD optimizers. 

Figure~\ref{fig:unet} shows the behavior of \textit{U-Net} architecture with \textit{CLMR} optimizer performing 2\% increase in dice index (in all three substructures as well as average) than its CRL optimizer counterpart. This proves that having a cyclic momentum rate can yield to better efficiency and accuracy than having a simple cyclic learning rate. The results on the test set comparing CLR and \textit{CLMR} optimizers in Table \ref{tab:results} support this conclusion too.

\begin{figure}[h]
\includegraphics[width=0.90\textwidth]{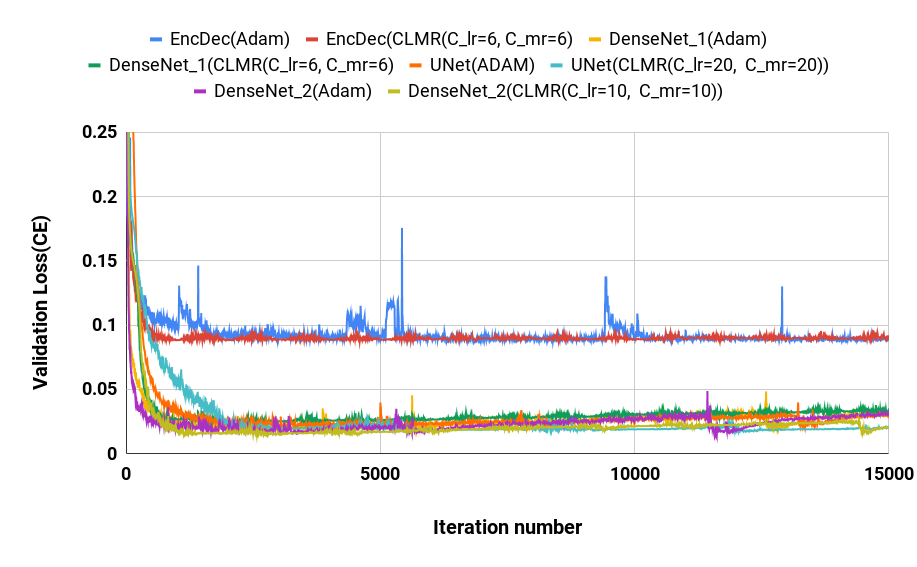}\label{fig:arch_diff_CE}
\includegraphics[width=0.90\textwidth]{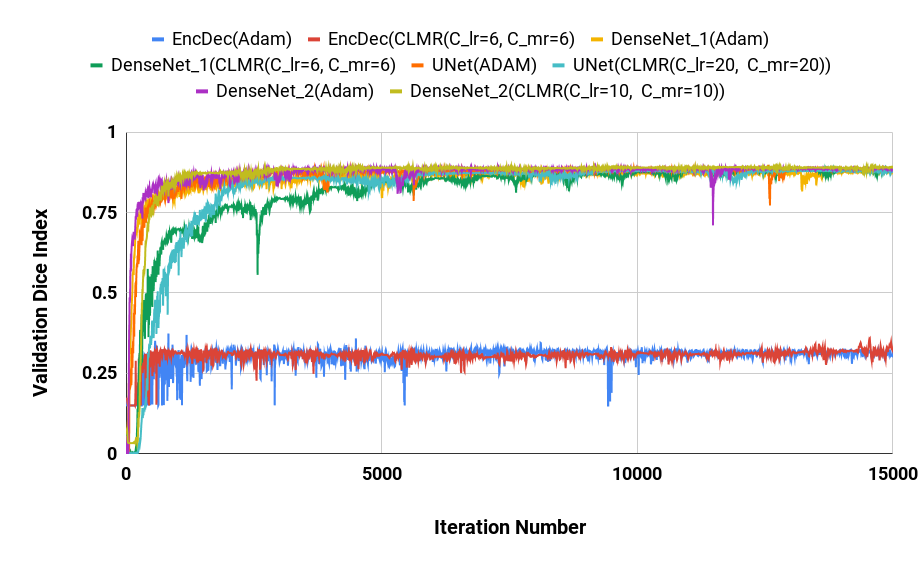}\label{fig:arch_diff_DI}
 \caption{Validation loss and dice index for four different architectures with \textit{ADAM} and \textit{CLMR} optimizers. (Upper) Cross Entropy loss in validation set for four different architectures. (Lower) Dice index in validation set for four different architectures (zoomed).  \label{fig:diff_arch}}
\end{figure}

Moreover, the curves of DI and CE among different architectures, trained by \textit{ADAM} and \textit{CLMR}, are demonstrated in Figure \ref{fig:diff_arch} a and b. Although the \textit{DenseNet}\_2 has less parameters in comparison with other architectures, it gets better results than the other architectures. These curves reveals some other important points about using different architectures: first, for all different architectures, the proposed \textit{CLMR} optimizer works better than \textit{ADAM} optimizer, indicating the power of proposed cyclic optimizer. Second, \textit{DenseNet} architectures are getting better results than \textit{U-Net} and Enc\_Dec architectures, which are highly over-parameterized architectures than \textit{DenseNet} and their saturation can be linked to this information too. Third, comparison between curves of \textit{DenseNet}\_1 and \textit{DenseNet}\_2 shows that having a higher GR (growth rate) in dense connections is more important than having dense block with high number of parameters. Since, \textit{DenseNet}\_2, with GR=24 reached better results in comparison with \textit{DenseNet}\_1 with twice of number of parameters in end of each dense block in comparison to \textit{DenseNet}\_2 and GR=16. These results are supported by the dice metric obtained from test data and are mentioned in Table \ref{tab:results}. 

\begin{figure}[h]
\centering
\includegraphics[width=1\textwidth]{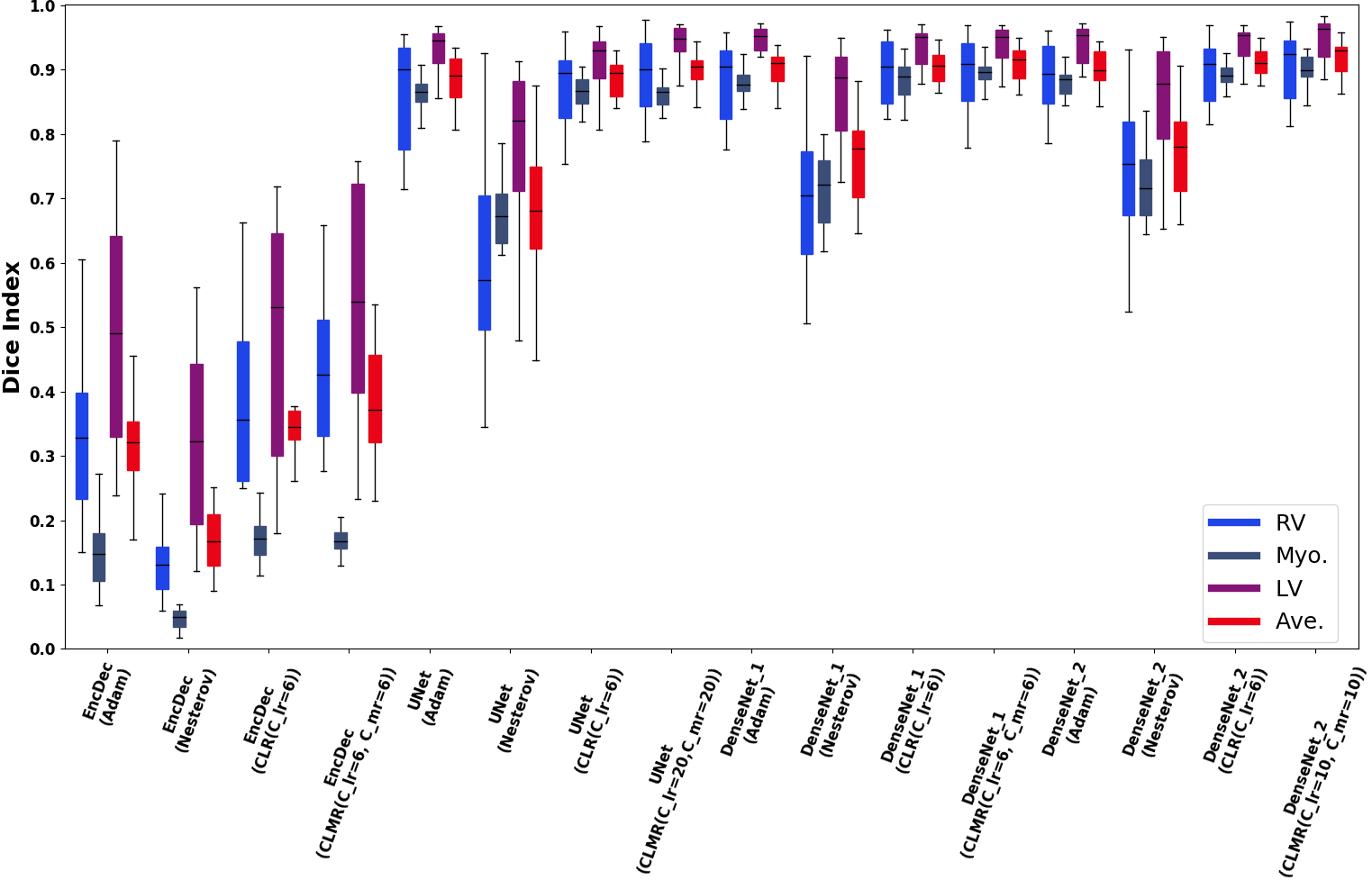}
\caption{Box plots for DI in test data set for RV, Myo., LV and also average of them (Ave).\label{fig:bp}}
\end{figure}

\begin{figure*}[ht!]
\centering
\includegraphics[width=0.9\textwidth]{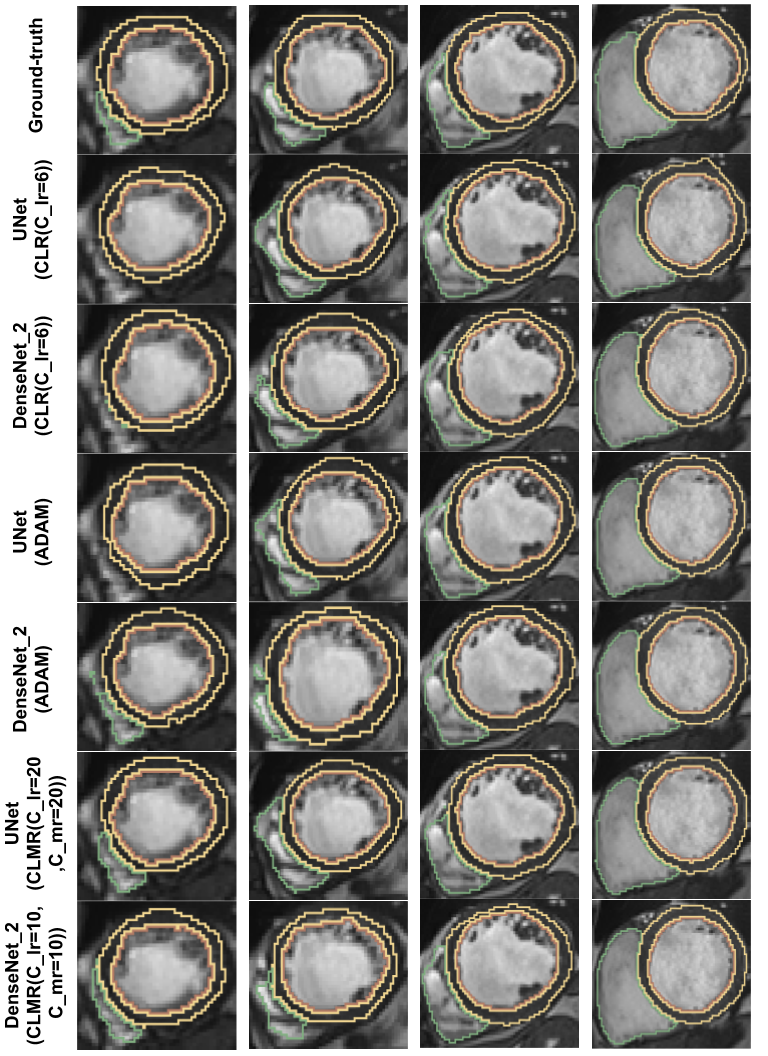}
\caption{Qualitative results for ground-truth and different methods for same subject in end-diastole from Apex to Base for four slices (from right to left). Green, yellow, and brown contours are showing RV, myo, and LV, respectively. \label{fig:opt_visual}}
\end{figure*}

\begin{figure*}[ht!]
\centering
\includegraphics[width=0.9\textwidth]{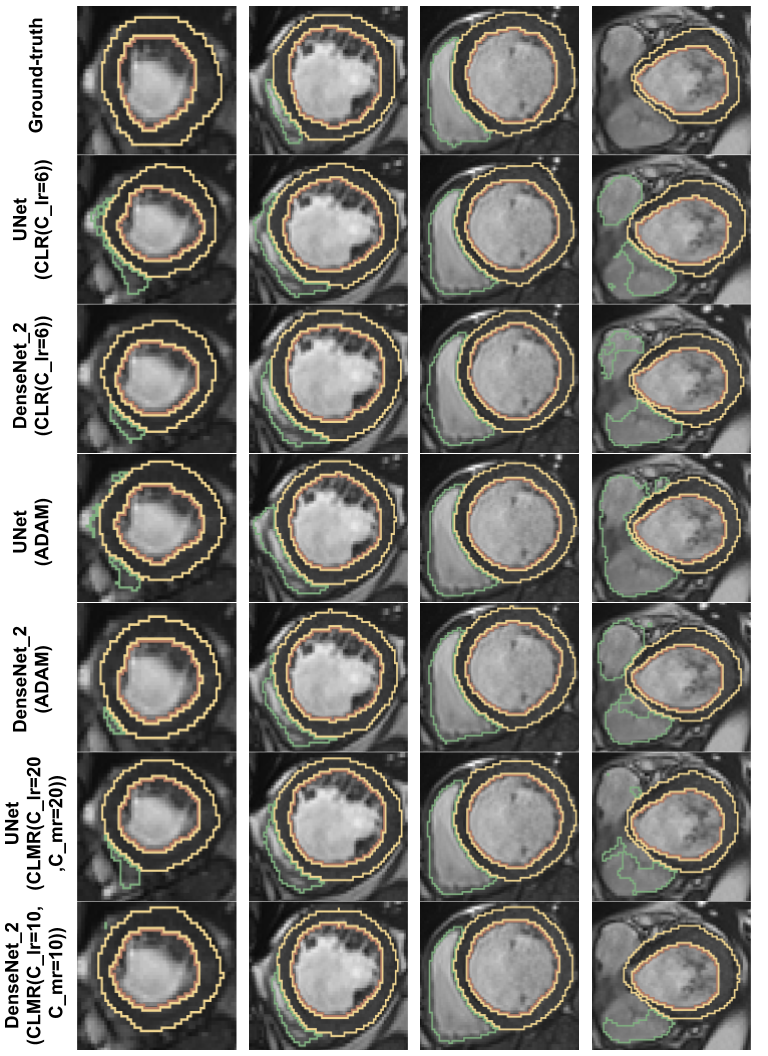}
\caption{Qualitative results for ground-truth and different methods for same subject in end-systole from Apex to Base for four slices (from right to left). Green, yellow, and brown contours are showing RV, myo, and LV, respectively. \label{fig:opt_visual_es}}
\end{figure*}

Finally, the DI on test data with online evaluation for different architectures with different optimizers are summarized in Table \ref{tab:results}. In order to have a better comparison, the box plot of all methods are drawn in Figure \ref{fig:bp}. As the figure shows, the dice statistic obtained from \textit{CLMR} is better than other optimizers most of the time in addition to its superior efficiency. In addition, qualitative results for different methods are shown in Figures \ref{fig:opt_visual} and  \ref{fig:opt_visual_es}: the contours for RV, Myo., and LV in ED for different methods and architectures and also ground-truth across four slices from Apex to Base. Usually, segmentation of RV near the Apex is harder than others because RV is almost vanishing at this point. As a result, some methods may not even detect the RV at slices near the Apex. Figure \ref{fig:opt_visual_es} shows the contours for RV, Myo., and LV in ES for different methods and architectures and also ground-truth across four slices from Apex to Base. Since at ES heart is at minimum volume; thus, it is more difficult to segment substructures. The contours generated with \textit{DenseNet}\_2 method is more similar to ground-truth in both ED and ES, which shows the generalizability of the  proposed method with an efficient architecture choice.  

\begin{table}
\caption{DI in the test data set with online evaluation.\\ \label{tab:results}}
\begin{tabular}{ll|l|l|l|l|}
\cline{3-6}
 &  & \textbf{Adam} & \textbf{Nesterov} & \textbf{CLR} & \textbf{CLMR} \\ \hline
\multicolumn{1}{|l|}{\textbf{Enc\_Dec}} & \multirow{4}{*}{\textit{RV}} & 0.3272 & 0.1309 & 0.3833 &  \textbf{0.4336} \\ \cline{1-1} \cline{3-6} 
\multicolumn{1}{|l|}{\textbf{U-Net}} &  & 0.8574 & 0.5968 & 0.8618 &  \textbf{0.8820} \\ \cline{1-1} \cline{3-6} 
\multicolumn{1}{|l|}{\textbf{DenseNet\_1}} &  & 0.8802 & 0.6936 &  \textbf{0.8961} & 0.8957 \\ \cline{1-1} \cline{3-6} 
\multicolumn{1}{|l|}{\textbf{DenseNet\_2}} &  & 0.8781 & 0.7232 & 0.8910 &  \textbf{0.9049} \\ \hline
\multicolumn{1}{|l|}{\textbf{Enc\_Dec}} & \multirow{4}{*}{\textit{Myo}} & 0.1473 & 0.1492 &  \textbf{0.1692} & 0.1686 \\ \cline{1-1} \cline{3-6} 
\multicolumn{1}{|l|}{\textbf{U-Net}} &  & 0.8628 & 0.6486 & 0.8588 &  \textbf{0.8631} \\ \cline{1-1} \cline{3-6} 
\multicolumn{1}{|l|}{\textbf{DenseNet\_1}} &  & 0.8787 & 0.7170 & 0.8834 &  \textbf{0.8960} \\ \cline{1-1} \cline{3-6} 
\multicolumn{1}{|l|}{\textbf{DenseNet\_2}} &  & 0.8796 & 0.7196 & 0.8904 &  \textbf{0.8999} \\ \hline
\multicolumn{1}{|l|}{\textbf{Enc\_Dec}} & \multirow{4}{*}{\textit{LV}} & 0.4950 & 0.3260 & 0.4972 &  \textbf{0.5418} \\ \cline{1-1} \cline{3-6} 
\multicolumn{1}{|l|}{\textbf{U-Net}} &  & 0.9238 & 0.7670 & 0.8936 &  \textbf{0.9360} \\ \cline{1-1} \cline{3-6} 
\multicolumn{1}{|l|}{\textbf{DenseNet\_1}} &  & 0.9376 & 0.8465 & 0.9351 &  \textbf{0.9393} \\ \cline{1-1} \cline{3-6} 
\multicolumn{1}{|l|}{\textbf{DenseNet\_2}} &  & 0.9196 & 0.8449 & 0.9378 &  \textbf{0.9478} \\ \hline
\multicolumn{1}{|l|}{\textbf{Enc\_Dec}} & \multirow{4}{*}{\textit{Ave.}} & 0.3232 & 0.1687 & 0.3499 &  \textbf{0.3814} \\ \cline{1-1} \cline{3-6} 
\multicolumn{1}{|l|}{\textbf{U-Net}} &  & 0.8813 & 0.6708 & 0.8714 &  \textbf{0.8937} \\ \cline{1-1} \cline{3-6} 
\multicolumn{1}{|l|}{\textbf{DenseNet\_1}} &  & 0.8988 & 0.7524 & 0.9049 &  \textbf{0.9103} \\ \cline{1-1} \cline{3-6} 
\multicolumn{1}{|l|}{\textbf{DenseNet\_2}} &  & 0.8924 & 0.7626 & 0.9064 & \textbf{0.9176} \\ \hline
\end{tabular}
\end{table}

\section{Discussions and Conclusions}
We proposed a new cyclic optimization method (\textit{CLMR}) to address the efficiency and accuracy problems in deep learning based medical image segmentation. We hypothesized that having a cyclic learning/momentum function can yield  better generalization in comparison to adaptive optimizers. We showed that CMLR is significantly better than adaptive optimizers by considering momentum changes inside the Nesterov optimizer as a cyclic function. Finding the parameters of these cyclic functions are complicated due to the correlation existing between LR and MR function. Thus, we formulated both LR and MR functions and we suggested a method to find the parameters of these cyclic functions with reasonable computational cost. 

Our proposed method is just a beginning of a new generation of optimizers which can generalize better than adaptive ones. One of the challenges in designing such optimizers is to set up the parameters of cyclic functions which need further investigation in a broad sense. One can learn these parameters with a neural network or reinforcement learning in an efficient manner: i.e., the $max_{lr}$, $min_{lr}$, $max_{,r}$, $min_{mr}$, $C_{lr}$, and $C_{mr}$ can be learned by an policy gradient reinforcement learning approach. In this study, our focus was only on supervised learning methods. However, proposed method can be generalized to semi-supervised or self-supervised methods as well. This is outside the scope of the current paper and can be thought of as a follow-up to what we proposed here.

In our study, our focus was in a particular clinical imaging problem: segmenting cardiac MRI scans. We assessed the optimization problem with single and multi-object settings. One may consider different imaging modalities and with different, and perhaps with newer, architectures to explore the architecture choices versus optimization functions. We believe that, based on our comparative studies, the architecture choice can affect the segmentation results such that more complex architectures require optimization algorithms to be selected wisely. 

The choice of optimization algorithm can depend on the specific characteristics of the dataset and the model being trained, as well as the computational resources available. Therefore, our results may not be generalizable to every situation in medical image analysis tasks. For instance, if the medical data is noisy or uncertain, it may be more difficult for the model to accurately predict the labels. This can make the optimization process more sensitive to the choice of optimization algorithm and may require the use of regularization techniques to prevent overfitting. For another example, if the dataset is highly imbalanced, with many more examples of one class than the other, it may be more difficult for the model to accurately predict the minority class. This can make the optimization process more challenging and may require the use of techniques such as class weighting or oversampling to improve the performance of the model. Last, but not least, if the dataset has a large number of features or the features are highly correlated, it may be more difficult to find a good set of weights and biases that accurately model the data. This can make the optimization process more challenging and may require the use of more advanced optimization algorithms.

Our study has some other limitations too. The use of second-order optimization methods are in high demands recently. However, it was not our focus on such methods due to their high burden in computational cost. Second-order optimization methods, which take into account the curvature of the loss function, have shown promising results in a variety of deep learning applications. These methods can be more computationally expensive than first-order methods, which only consider the gradient of the loss function, but may be more effective in certain situations. Further, we focused on the segmentation problem with traditional deep network architectures while reinforcement learning and generative models can require development of new algorithms tailored to specific types of problem.

\subsection{Disclosures}
No Conflict of Interest.

\subsection{Acknowledgments}
This project is supported by NIH funding: R01-CA246704, R01-CA240639, U01-DK127384-02S1, R03-EB032943-02, and R15-EB030356. 

\subsection{Data, Materials, and Code Availability}
Data is available under MICCAI 2017 ACDC challenge.


\bibliography{refs}   
\bibliographystyle{spiejour}   


\vspace{1ex}
\noindent Biographies and photographs of the other authors are not available.

\listoffigures
\listoftables

\end{spacing}
\end{document}